# Eclipsing binary V658 Carinae (HD 92406) containing a shell star and a hot low-mass star in the post-Algol stage

Norbert Hauck

Bundesdeutsche Arbeitsgemeinschaft für Veränderliche Sterne e.V. (BAV), Minsterdamm 90, 12169 Berlin; hnhauck@yahoo.com

**Abstract:** *Further studies of the binary V658 Car (HD 92406) have now confirmed its unique main features, i.e. a shell star primary component being eclipsed by a hot low-mass star. Obviously, this binary is in its post-Algol evolutionary stage, and containing a rapidly rotating primary star and a contracting hot companion star. The analysis of old radial-velocity data indicating a primary star of lower mass has now resulted in amended parameters for the binary. According to stellar models the primary star should have a mass of 2.16 ± 0.14 Msun and a radius of 1.64 ± 0.04 Rsun for a Teff of 9700 ± 500 K based on known spectral type A0p. For the secondary star 1.46 ± 0.04 Rsun and 12750 ± 700 K Teff have been derived and are fitting well to an existing evolutionary track for a contracting white dwarf precursor of 0.28 Msun. Our photometric results have confirmed the concave shape of the large decretion disk surrounding the primary star, as predicted by existing models.*

V658 Car (HD 92406) is the first binary showing eclipses of a shell star and its surrounding equatorial gas disk by the companion star. This hot low-mass companion obviously is the former mass-transferring loser in the middle of its post-Algol contracting stage, what equally has never been seen before. Both discoveries have already been presented and discussed in an earlier paper of the author [1]. The analysis of new photometric and spectroscopic data as well as old radial-velocity data has now confirmed these findings. Nevertheless, the parameters of the binary components had to be amended.

The photometry in UBVIc and H alpha with a remote controlled 0.5m-reflector telescope of an observatory in Siding Spring (SSO), Australia, has been continued. From the ROAD observatory (0.4m reflector) at San Pedro, Chile, additional photometry of the central star eclipses in passbands B and Ic has been contributed.

Figures 1, 2, 3 and 4 show the light curves in U, B, Ic and H alpha, respectively, folded over the orbital period. As indicated in the to-scale sketch of Fig. 7, the first and last contact phase of the outer primary eclipse, when the hotter secondary star 2 is covered by the large gas disk of primary star 1, are in orbital phases 0.905 and 0.115. Hence a disk of an assumed circular shape would be in a slightly eccentric position to star 1 as shown in Fig. 7. For the first time, additional minima have been discovered at phases 0.965 and 0.035, i.e. just before and after the primary central star eclipse (see e.g. Fig. 1). This feature is regarded as a convincing indication of the presence of a concavely shaped gas disk surrounding star 1 in accordance with the existing standard model for such a decretion disk (see T. Rivinius et al. [2]). Obviously, in our edge-on view of the binary, the light of star 2 is being increasingly dimmed as a function of its covered distance across the thicker, outer part of the gas disk, when it approaches phase 0. However, when the inner disk becomes thinner than the size of star 2, the gas column in front of the star loses total absorption power towards our line of sight, and star 2 will become brighter again. Thereby, we have got here a new and direct optical proof of the correctness of the existing concavely shaped model for decretion disks. Of course, the visible brightening ends at central star eclipse, i.e. between phase 0.992 and 0.008.

The extended outer secondary minimum from about phase 0.35 to 0.65 is shallower than the outer primary minimum in all observed passbands (see Fig. 1 to 4) and has no simple explanation. A quantitative study of V658 Car shows that the eclipse of the large luminous disk by much smaller star 2 alone does not fit to the duration and depth of this minimum. Adding a large accretion disk to star 2 does no longer appear to be acceptable, since no sufficient loss of light has been found in the UBVIc light curves just before and after the central secondary eclipse (phase 0.492 to 0.508). The minor drop of light from about 0.025 mag in U to about 0.04 mag in Ic from phase 0.48 to 0.49 is attributed to star 2 partially eclipsing the bright so-called 'pseudo-photosphere', which normally is extended to about twice the radius of the gas-losing primary star (see cited reference [2]). Reference [2] also contains a discussion indicating that our outer secondary minimum might be created by local densifications in the gas disk caused by tidal interaction with star 2.

The light curve in passband H alpha is also shown (see Fig. 4), but is more difficult to understand. Apparently, the total light in H alpha is phase-locked as well and the result of variable emission and absorption. For a deeper understanding we have to look into the H-alpha spectra, which have been obtained by B. Heathcote [3] and are available online in the BESS database. Although their interpretation is not the specific subject of the present paper, these spectra show H-alpha emission-line profiles being typical for shell stars, and a central absorption increasing towards orbital phases 0 and 0.5.

Radial-velocity data of V658 Car obtained in an earlier study by F. Gieseking [4] have now been reexamined taking into account the results of our first paper [1]. After elimination of the first 3 data points of the second data set (data points nos. 1 and 3 had already been rejected in reference [4]), a much better $\sigma_{FIT}$ of 6.5 km/s (instead of 10.0 km/s) has been achieved with help of the *Binary maker 3* software. This under the logical assumption, that 4 of the remaining 14 measurements should be attributed to the bright and low-mass secondary component of the binary (see Fig. 6). Surprisingly, our fit parameters indicated a much lower mass of the primary component than estimated in our first study [1]. It soon became clear that this has been caused simply by a wrong relationship between binary components and known, but misleading spectral classes. Apparently, the shell feature and star 1 should be rather attributed to spectral type A, and spectral type B linked to our secondary component. Although shell stars are normally known as Be shell stars, they can also sometimes appear in spectral class A (or even F) (see ref. [2]). Therefore, primary star 1 has now been shifted from B5Vp shell (from ref. [11]) to spectral type A0p (from F. Gieseking [5]) corresponding to a $T_{eff}$ of about 9700 ± 500 K. From RAVE data release 5 [6] we get for V658 Car a $T_{eff}$ of 6144 ± 106 K, which should be the mean temperature of the gas disk usually having about 60% of the $T_{eff}$ of its central star, according to ref. [2].

Computed light curves have been fitted to our new UBVIc photometric data of the central star eclipses with *Binary maker 3* in a similar way as already described in our former paper [1], and achieved a $\sigma_{FIT}$ of 4 resp. 5 mmag in passband U (see Fig. 5). The set of parameters fitted in U has been confirmed in passbands BVIc, however, for the $T_{eff}$ of star 2 a significantly higher value of 15300 K has been obtained in Ic, compared to a weighted mean value of 12750 K in UBV. This deviation is regarded as a consequence of peculiar absorption/emission behavior of the decretion disk, which, apparently, is distorting here the normal $T_{eff}$ – luminosity relation. The effect can be compensated by adjusting disk light $l_3$ in the Ic-model.

The ratio $F_1$ of the rotational/orbital period of star 1 cannot be found easily. All oscillations detected in ASAS-3's photometric data of the maximum appear to be observational aliases rather than true signals of pulsations or rotation of the star. Therefore, the well determined rotation frequencies of slowly pulsating Be (SPBe) stars from the *MOST* mission as shown in Fig. 15 of C. Cameron et al. [7] have been used as a starting point for modeling $F_1$. Extrapolation to our $T_{eff}$ and optimization in our fit gave a final $F_1$ value of 87.7 ($F_2$ = 1 adopted). The ratio w of equatorial / critical rotation velocity of star 1 then gives 237 km/s / 490 km/s, i.e. about 0.48. This is significantly below the expectations of w = 0.77 ± 0.08 for a such a shell star, according to the review of T. Rivinius et al. [2], and might be interpreted as an indication of differential, more rapid rotation in the equatorial zone of our primary star.

The stellar model with rotation from S. Ekström et al. [8] gives for solar metallicity Z=0.014, $T_{eff}$ 9700 K and our $R_1$/a (radius/separation) ratio of 0.0287 the best fit for star 1 onto the ZAMS (zero age main sequence) corresponding to 2.16 solar masses (Msun) and 1.64 solar radii (Rsun). Simultaneously, our star 2 ($T_{eff}$ 12750 K and 1.46 Rsun) is on the evolutionary track of a proto-helium white dwarf of 0.28 Msun and has an age of not more than 7 Myr calculated from Roche-lobe detachment, according to Fig. 1 and A1 of A.G. Istrate et al. [9]. Moreover, the same $M_2$ has been obtained by our radial-velocity modeling for star 2, if star 1's mass is fixed to 2.16 Msun. The $M_2$ of 0.28 Msun also fits to the white dwarf remnant mass – final orbital period relation (for stable Roche-lobe overflow) given in equation (29) and Fig. 6 of Carter et al. [10].

Our results are listed in tables 1 and 2. The error margins are based on the estimated error margin of 500 K for star 1's $T_{eff}$ of 9700 K. The orbital period appears to be constant over the last 17 years of observation.

The discovery and confirmation of the main features of V658 Car should motivate better equipped astronomers to further studies. This unique binary offers new possibilities in the research of rapidly rotating stars, decretion disks and post-Algol evolution.

**Table 1: Parameters of binary system V658 Car**

| | | |
|---|---|---|
| Epoch [HJD] | 2452786.438(2) | mid Prim. Min.; from ASAS + new data |
| Period [days] | 32.1854(1) | constant for JD 2451900 – 2458145 |
| Total light in U/B/V [mag] | 8.64/9.08/9.11 | from Buscombe catalogue [11] |
| Minimum duration [hours] | 12.4 | partial eclipses |
| Orbital inclination i [deg] | 88.72 | (-0.09/+0.18) |
| Orbital radius a [R$_\odot$] | 57.27 ± 1.17 | for R$_\odot$ = 696342 km; circular orbit |
| RV of K$_1$ / K$_2$ [km/s] | 10.35 / 79.6 | from RV modeling (M$_1$ fixed to 2.16 M$_\odot$) |
| RV barycenter [km/s] | 28.6 ± 2.9 | (RV = Radial velocity) |
| Mass ratio q (M$_2$/M$_1$) | 0.13 ± 0.01 | from the masses of table 2 |
| Distance [pc] | 589 ± 38 | calculated for A$_v$ = 0 |

**Table 2: Parameters of components of V658 Car**

| Parameter | Primary star | Secondary star | Disk |
|---|---|---|---|
| Spectral type (amended) | A0p(e shell) | B | |
| Temperature T$_{eff}$ [K] | 9700 ± 500 | 12750 ± 700 | 6144 ± 106 |
| Radius R (volume) [R$_\odot$] | 1.64 ± 0.04 | 1.46 ± 0.04 | 33.6 ± ≥0.7 |
| Luminosity (bol.) [L$_\odot$] | 21.5 ± 5.6 | 50.7 ± 13.2 | |
| Brightness (abs.) [VMag] | 1.59 | 1.31 | |
| U-light fraction at Max. | 0.173 dimmed | 0.534 | 0.293 |
| B-light fraction at Max. | 0.202 dimmed | 0.445 | 0.353 |
| V-light fraction at Max. | 0.183 dimmed | 0.382 | 0.435 |
| Ic-light fraction at Max. | 0.142 dimmed | 0.347 | 0.511 |
| Mass [M$_\odot$] | 2.16 ± 0.14 | 0.28 ± 0.01 | |

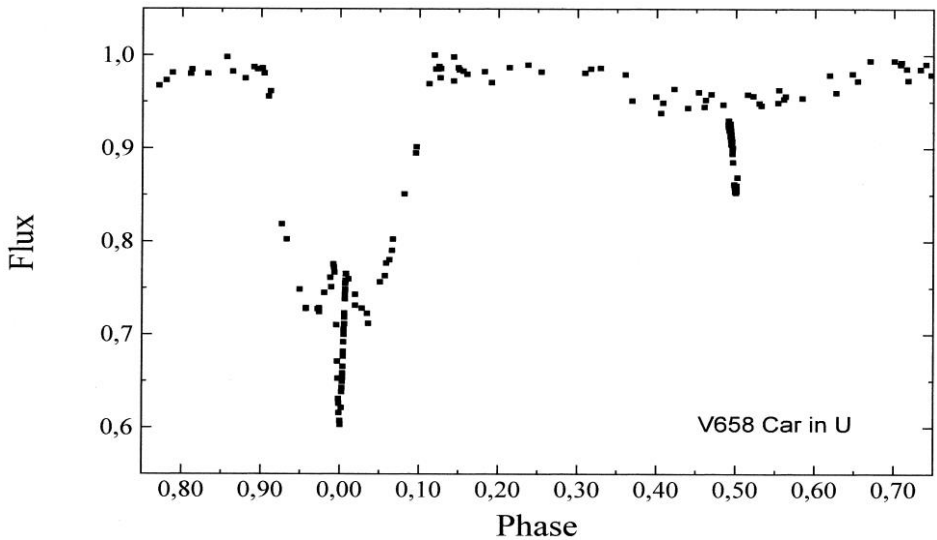

**Fig. 1:** U-Flux – phase diagram of V658 Car from 167 data points from SSO

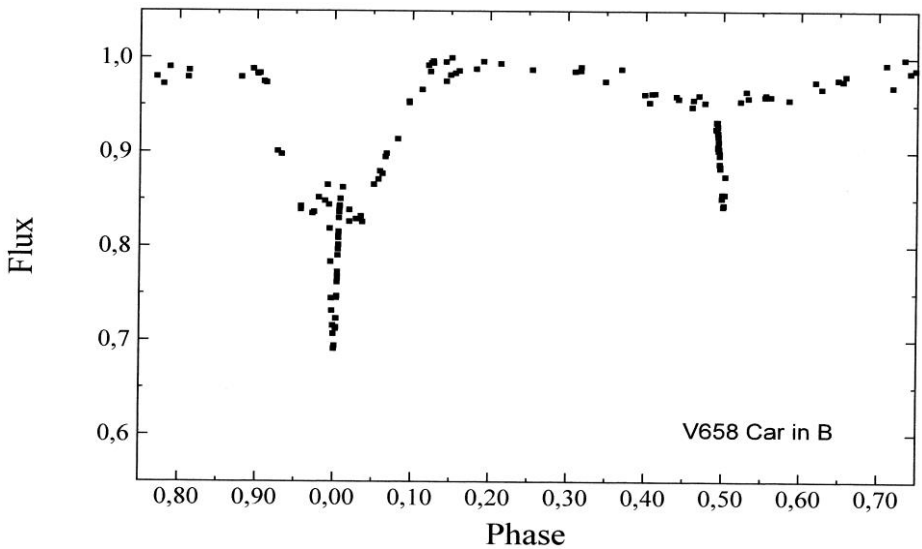

**Fig. 2:** B-Flux – phase diagram of V658 Car from 139 data points from SSO

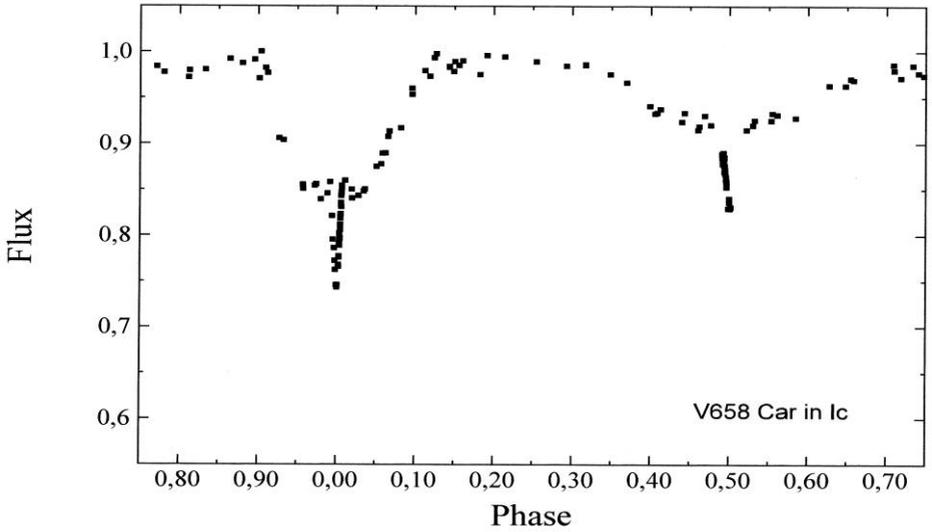

**Fig. 3:** Ic-Flux – phase diagram of V658 Car from 138 data points from SSO

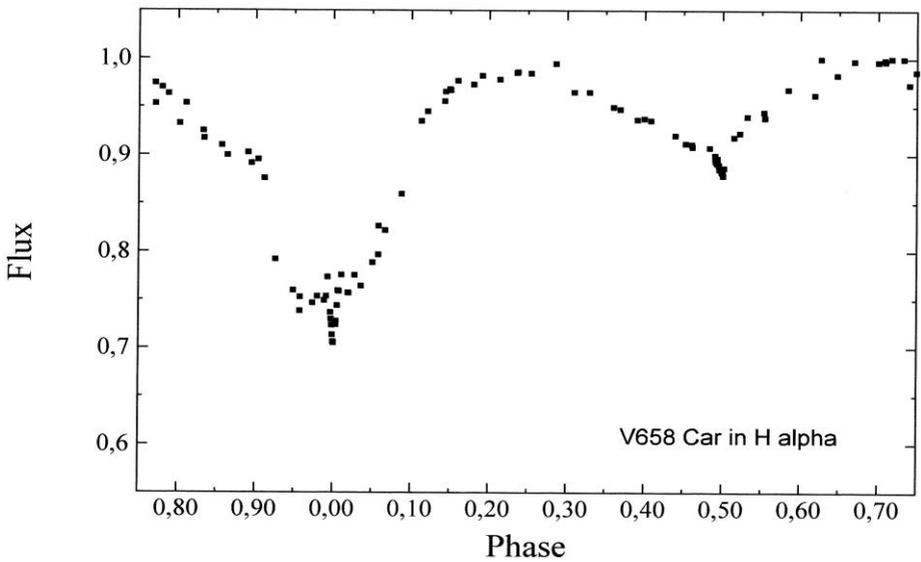

**Fig. 4:** H-alpha Flux – phase diagram of V658 Car from 104 data points from SSO

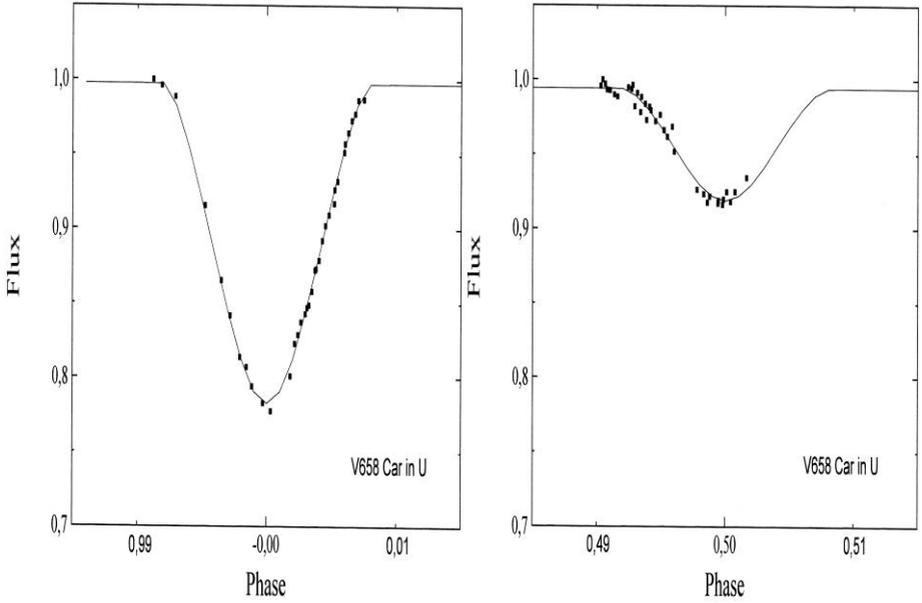

**Fig. 5:** Computed light curves for U data of the central star eclipses of V658 Car

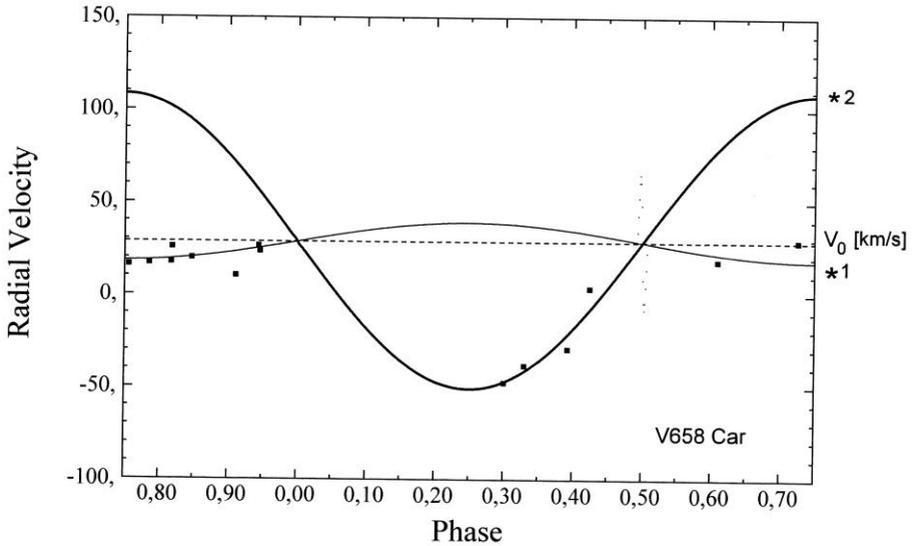

**Fig. 6:** Radial velocities [km/s] of the binary V658 Car with 14 data points from [4]

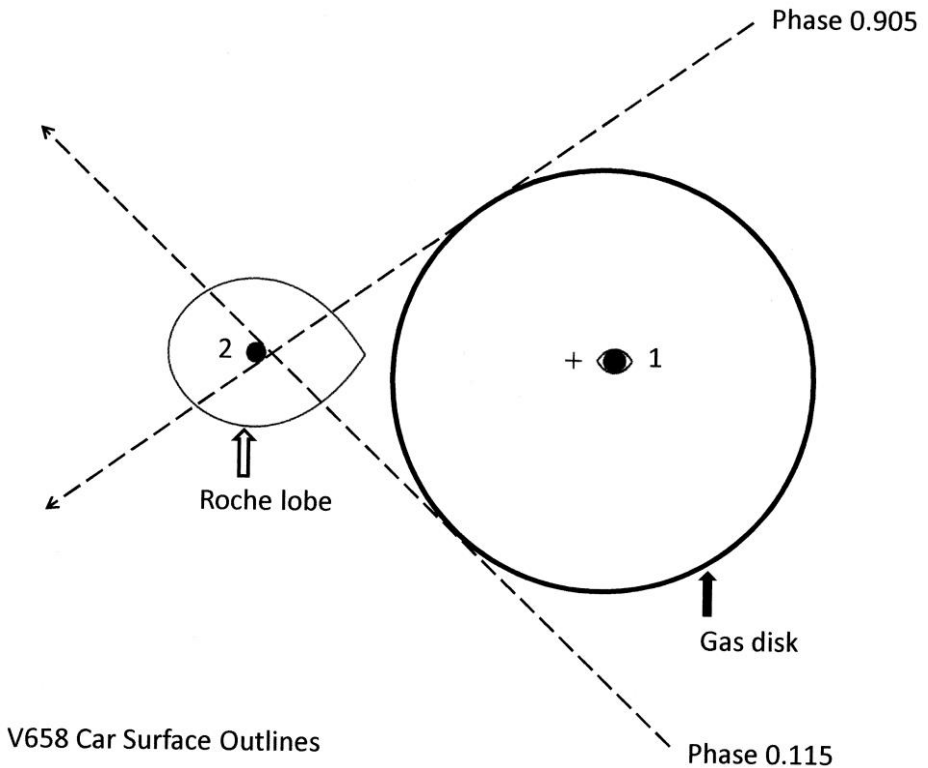

**Fig. 7:** A to-scale sketch of V658 Car with the two stars, their critical surface outlines and the barycenter (+) (in cross-sectional view). The slightly eccentric positioned equatorial gas disk of the primary star and the direction of our line of sight at the first and last contact phase of the outer primary eclipse have been added in face-on view. Rapidly rotating primary star 1 nearly fills out its critical limiting lobe.

**Acknowledgements:**
I am grateful for the helpful comments of Dr. T. Rivinius, ESO, Chile, to my earlier paper on V658 Car and the H-alpha spectra of this binary, which have been kindly taken by B. Heathcote, Australia. Many thanks also to Dr. F.-J. Hambsch, Belgium, for the additional photometric data collected with his ROAD observatory in Chile. This research has made use of the Simbad and VizieR databases operated at the Centre de Données astronomiques de Strasbourg, France, http://cdsarc.u-strasbg.fr/ and the All Sky Automated Survey ASAS database, http://www.astrouw.edu.pl/asas/